\documentclass[3p]{elsarticle}
\usepackage{stfloats}

\usepackage[utf8x]{inputenc}
\usepackage{amsmath}
\usepackage{amsfonts}
\usepackage{amssymb}
\usepackage{graphicx}
\usepackage{subfig}
\usepackage{natbib}
\usepackage{hyperref}
\usepackage{todonotes}

\graphicspath{{figs/}}
\journal{tbd}

\begin{document}
\title{Machine-learning–enabled interpretation of tribological deformation patterns in large-scale MD data}
% \title {Machine-Learning-Based Identification of Deformation Patterns in Molecular Dynamics Simulations via Learned Feature Representations}
% \title {Automated Identification of Deformation Patterns in Molecular Dynamics Simulations via Learned Feature Representations}

% \author[KIT,µTC]{A.~Kashyap} % ORCID:
% \author[ac2t,TUWien]{H.~J.~Ehrich}
% \author[ac2t]{M.~Rodr\'iguez Ripoll}
% \author[ac2t,TUWien]{S.~J.~Eder\corref{cor1}} % 0000-0002-2902-3076
% \ead{stefan.j.eder@tuwien.ac.at}
% \cortext[cor1]{Corresponding author}
% \author[KIT,µTC]{J.~Schneider} % ORCID:

% \address[KIT]{Karlsruhe Institute of Technology, Karlsruhe, Germany}
% \address[µTC]{$\mu$TC, Karlsruhe, Germany}
% \address[ac2t]{AC2T research GmbH, Viktor-Kaplan-Stra\ss e 2/C, 2700 Wiener Neustadt, Austria}
% \address[TUWien]{Institute for Engineering Design and Product Development, TU Wien, Leh\'argasse 6 -- Objekt 7, 1060 Vienna, Austria}

\author[1,2]{H. J. Ehrich}%[role=Co-ordinator, suffix=,  ]
% \credit{Methodology, Software, Visualization, Writing - Original Draft, Data Curation}

\author[3]{M. C. May}

\author[1,2]{S. J. Eder\corref{cor1}}%[orcid=0000-0001-7511-2910]
                        % auid=000,bioid=1,
                        % prefix=,
                        % role=Researcher,
                        % 
% \cormark[1]
\ead{stefan.j.eder@tuwien.ac.at}
% \credit{Conceptualization, Supervision, Writing - Review \& Editing, Project administration, Data Curation}

% affiliations

\affiliation[1]{organization={Institute for Engineering Design and Product Development, TU Wien},
    addressline={Lehárgasse 6 – Objekt 7}, 
    city={1060 Vienna},
    country={Austria}}
% \affiliation[2]{organization={Institute for Applied Materials (IAM), Karlsruhe Institute of Technology (KIT)},
%     addressline={Kaiserstraße 12}, 
%     city={76131 Karlsruhe}, 
%     country={Germany}}
% \affiliation[3]{organization={MicroTribology Center (µTC)},
%     addressline={Straße am Forum 5},
%     city={76131 Karlsruhe}, 
%     country={Germany}}
\affiliation[2]{organization={AC2T research GmbH},
     addressline={Viktor-Kaplan-Straße 2/C}, 
    city={2700 Wiener Neustadt},
    country={Austria}} 
\affiliation[3]{organization={School of Mechanical \& Aerospace Engineering, Nanyang Technological University},
addressline={50 Nanyang Avenue},
 country={639798 Singapore}}
% Corresponding author text
\cortext[cor1]{Corresponding author}

\begin{abstract}
Molecular dynamics (MD) simulations have become indispensable for exploring tribological deformation patterns at the atomic scale. 
However, transforming the resulting high-dimensional data into interpretable deformation pattern maps remains a resource-intensive and largely manual process. 
In this work, we introduce a data-driven workflow that automates this interpretation step using unsupervised and supervised learning. 
Grain-orientation-colored computational tomograph pictures obtained from CuNi alloy simulations were first compressed through an autoencoder to a 32-dimensional global feature vector. 
Despite this strong compression, the reconstructed images retained the essential microstructural motifs: grain boundaries, stacking faults, twins, and partial lattice rotations—while omitting only the finest defects. 
The learned representations were then combined with simulation metadata (composition, load, time, temperature, and spatial position) to train a CNN--MLP model to predict the dominant deformation pattern.
The resulting model achieves a prediction accuracy of approximately 96\% on validation data. 
A refined evaluation strategy, in which an entire spatial region containing distinct grains was excluded from training, provides a more robust measure of generalization. 
The approach demonstrates that essential tribological deformation signatures can be automatically identified and classified from structural images using Machine Learning. 
This proof of concept constitutes a first step towards fully automated, data-driven construction of tribological mechanism maps and, ultimately, toward predictive modeling frameworks that may reduce the need for large-scale MD simulation campaigns.
\end{abstract}


\maketitle

\section{Introduction}

Friction and wear are major sources of energy loss and material degradation across engineering systems, contributing to roughly 23\% of global energy consumption being dissipated in tribological contacts \cite{holmberg_influence_2017}. 
Understanding how materials deform under tribological loading is therefore essential for improving durability and energy efficiency. 
At the atomic scale, molecular dynamics (MD) simulations can provide detailed insights into the microstructural evolution and deformation patterns under controlled variations in composition, pressure, temperature, and sliding conditions \cite{eder_unraveling_2020}. 
Such simulations have been used to study a broad range of metals and alloys, including Al \cite{dan_microstructure_2012, shi_effect_2021}, Ni \cite{dmitriev_molecular_2017}, Fe \cite{romero_coarse_2014, eder_thermostat_2017}, Mo \cite{hu_tensile_2023}, or Cu \cite{romero_coarse_2014, li_how_2015}, among others. 
Such works highlight the roles of dislocation activity, stacking faults, and twinning in plastic deformation \cite{shi_effect_2021, dong_molecular_2024}, as well as the strong influence of grain size and alloy composition on the tribological response \cite{hansen2004hall, schuh2002hardness, argibay_linking_2017}.

Besides the considerable computational expense of MD simulations,the significant manual effort required for interpreting large volumes of structural data is its major limitation. 
This challenge persists even when simulations avoid nanocrystalline artifacts associated with the Hall-Petch breakdown by considering sufficiently large grains \cite{eder_unraveling_2020, eder_interfacial_2018}. 
Existing automated tools, such as twin identification frameworks \cite{ehrich_automated_2024}, address isolated tasks but do not provide a general method for identifying dominant deformation patterns across diverse tribological conditions.

In parallel, machine learning (ML) has transformed many areas of materials research by enabling large-scale screening, structure–property prediction, and microstructure optimization \cite{zhou_big_2019, louie_discovering_2021, bishara_state---art_2023, yin2024ai}. 
ML approaches have been applied to predict mechanical properties of complex alloys \cite{zhang_molecular_2021}, optimize microstructures \cite{lin_neural_2023}, and analyze heterogeneous experimental tribological data \cite{yin_tribo-informatics_2023, paturi_role_2023}. 
Convolutional neural networks (CNNs) in particular offer powerful tools for interpreting high-dimensional image data and have been used to classify atomic environments \cite{fukuya_machine_2020}, extract grain boundary features \cite{gomberg_extracting_2017}, and interpret diffraction patterns \cite{kaufmann2020phase}. 
Although approaches using CNNs have attempted to classify wear mechanisms from scanning electron microscopy (SEM) images~\cite{sieberg2022wear, SIEBERG2023204725}, similar links to tribological molecular dynamics simulations remain largely unexplored.
Existing efforts primarily target isolated structural descriptors or mechanical property predictions \cite{jin_machine_2023, minkowski_predicting_2023}, and no available method performs automated, mechanism-level classification of deformation modes in large-scale polycrystalline MD data. 
A recent study by Chen et al. \cite{chen_detecting_2023} comes closest in scope but focuses on identifying structural criticalities rather than classifying deformation regimes.

In this work, we introduce a data-driven ML workflow for the automated interpretation of structural images from MD simulations of tribologically loaded polycrystals.
Grain-orientation-colored computational tomographs are first compressed using a deep autoencoder to explore low-dimensional structure representations.
Independently of this, a dual-branch CNN–MLP model is trained directly on the raw images together with simulation metadata to predict dominant deformation mechanisms.
Our results demonstrate that key tribological deformation signatures can be learned directly from structural images, offering an outlook toward automated construction of tribological deformation pattern maps and enabling more efficient use of atomistic simulations in materials discovery.

This paper is structured as follows: we first briefly introduce how computational materials tribology based on MD simulations has been carried out over the past years, motivating the necessity for more resource efficient methods.
We then walk the reader through our labeling process, the handling of the large amount of historic MD data, and an unsupervised learning approach.
At the heart of this contribution lies the architecture of a supervised prediction model, including a suitable strategy for splitting between training and validation data as well as an approach to assess the interpretability of the model's output.
We then apply the introduced methods to a consistent data set consisting of computational tomographs of tribologically loaded CuNi alloy surfaces plus accompanying metadata.
We discuss the ML model's performance and interpret its output, thus conducting sanity checks on the predictions.
Finally, a broader outlook includes possible avenues of further work to make such models a part of engineering tools of the future.

\section{Background and Motivation}
\subsection{Conventional Data Processing Workflow}

% \section{Background and Conventional Workflow}
% \subsection{Traditional Data Processing}

\begin{figure}[htbp]
\centering
\includegraphics[width=1.0\textwidth]{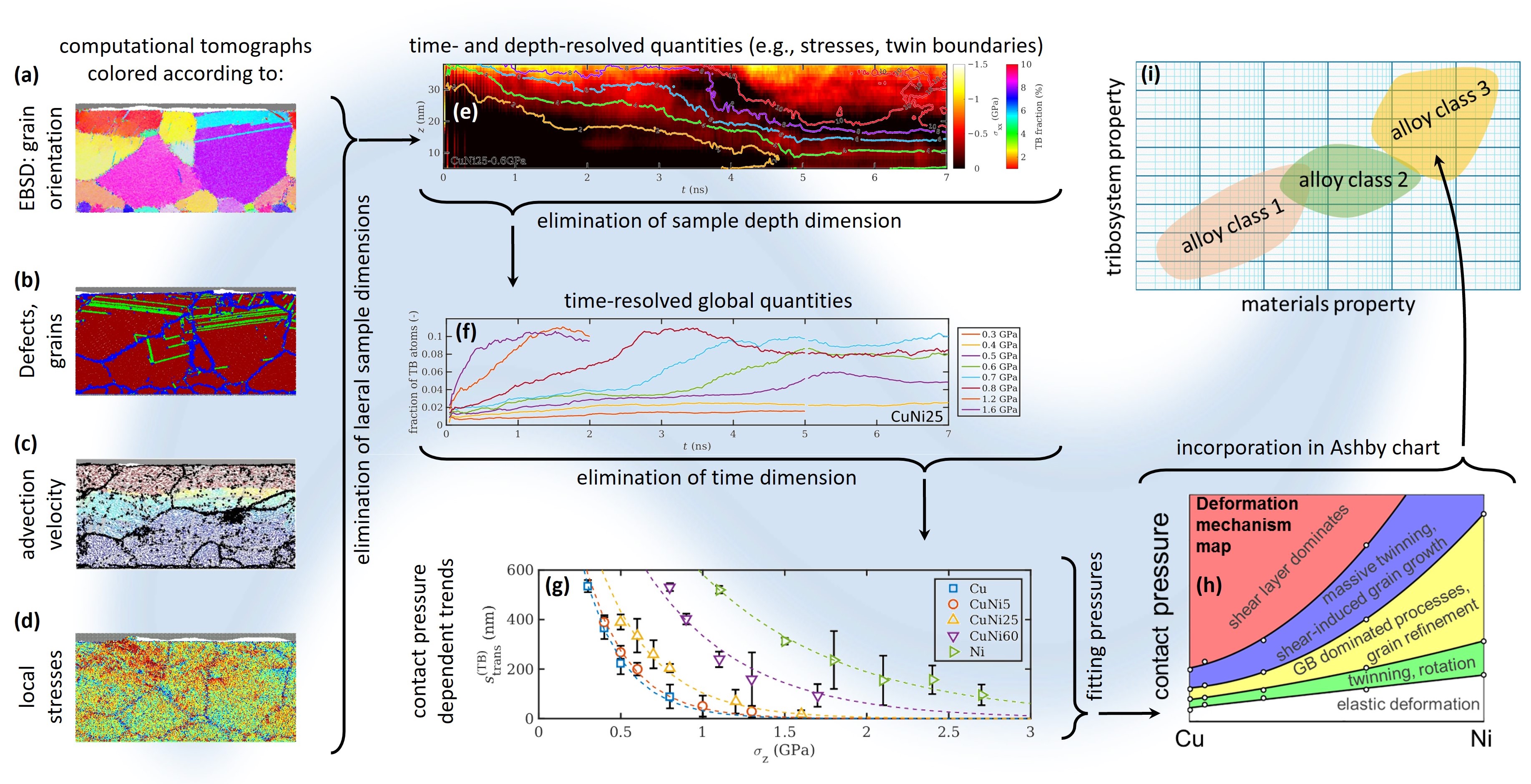}
\caption{Schematic of conventional MD data analysis. The data distillation workflow follows the tilde-shaped background from (a,b,c,d) to (i), based on the representative example of statistically evaluating the emergence of twin boundaries (TB) in the microstructure. A fully developed ML model could replace all the intermediate steps up to the deformation pattern map.}
\label{fig:conv_MD_data_analysis}
\end{figure}

Conventional MD analysis, which has been used in previous data evaluation~\cite{eder_unraveling_2020, eder_effect_2020, eder_does_2022} and can serve labeling and validation purposes for ML model construction and preparation, employs a multi-tiered data distillation process to derive robust trends, see Fig.~\ref{fig:conv_MD_data_analysis}.
In the left column of this figure, we show representative examples of computational tomographs through the 3D MD model, with the atoms colored by (a) grain orientation in electron backscatter diffraction (EBSD) standard, (b) lattice type, grain boundaries, and defects, (c) advection (drift) velocity to visualize shearing, and (d) local stresses.
As a first step in the data distillation process, these 3D data that are stored for each atom are averaged across the lateral system dimensions, revealing depth-resolved, time-dependent quantities of interest, as visualized in the heat map at the top of the middle column (e).
Further elimination of the sample depth and time dimensions leads to time-resolved global quantities (f) and contact pressure dependent trends (g), which can be fitted with characteristic pressures that mark the transition between deformation patterns (h).
As an outlook to the utility of such highly distilled data, we propose their incorporation into Ashby-style charts, as schematically shown in Fig.~\ref{fig:conv_MD_data_analysis}~(i), which link material properties with tribological properties.
This conventional approach accommodates the complexities of polycrystalline materials under tribological loading conditions and is guided by the underlying physics, resulting in this time-consuming procedure. Thus, substituting this approach with a well-trained ML model is highly relevant. The conventional approach can serve as the ground truth for training this ML model or to refine and validate said model based on newly generated MD data.

\subsection{Computational Cost}
Each MD simulation run of 7~ns of sliding for a 25m atom system typically produces some 200~GB of raw compressed data and requires 160k CPU-h for both computation and postprocessing.
There are approximately 60 data points available for 5 CuNi alloys at two temperatures and various contact pressures.
In addition, there are approximately 20 more points available for 4 CuNi alloys at various sliding speeds.
This leads to some 16~TB of raw data that took around 13m CPU-h to calculate, which corresponds to a wall time of approximately one year at a job size of 1440 CPU cores.
It therefore becomes clear that the computational expense for a reasonable parametric sweep for one single alloy class is considerable.
Another bottleneck is the interpretative step, which requires human resources to evaluate the visualizations and guide the data distillation
This motivates the search for automated methods that can on one hand reduce the amount of necessary calculations and on the other hand dramatically cut corners during the laborious interpretation.

\section{Methodology Overview}
\subsection{MD Data and System Description}

\begin{figure}[htbp]
    \centering
    \includegraphics[width=0.8\linewidth]{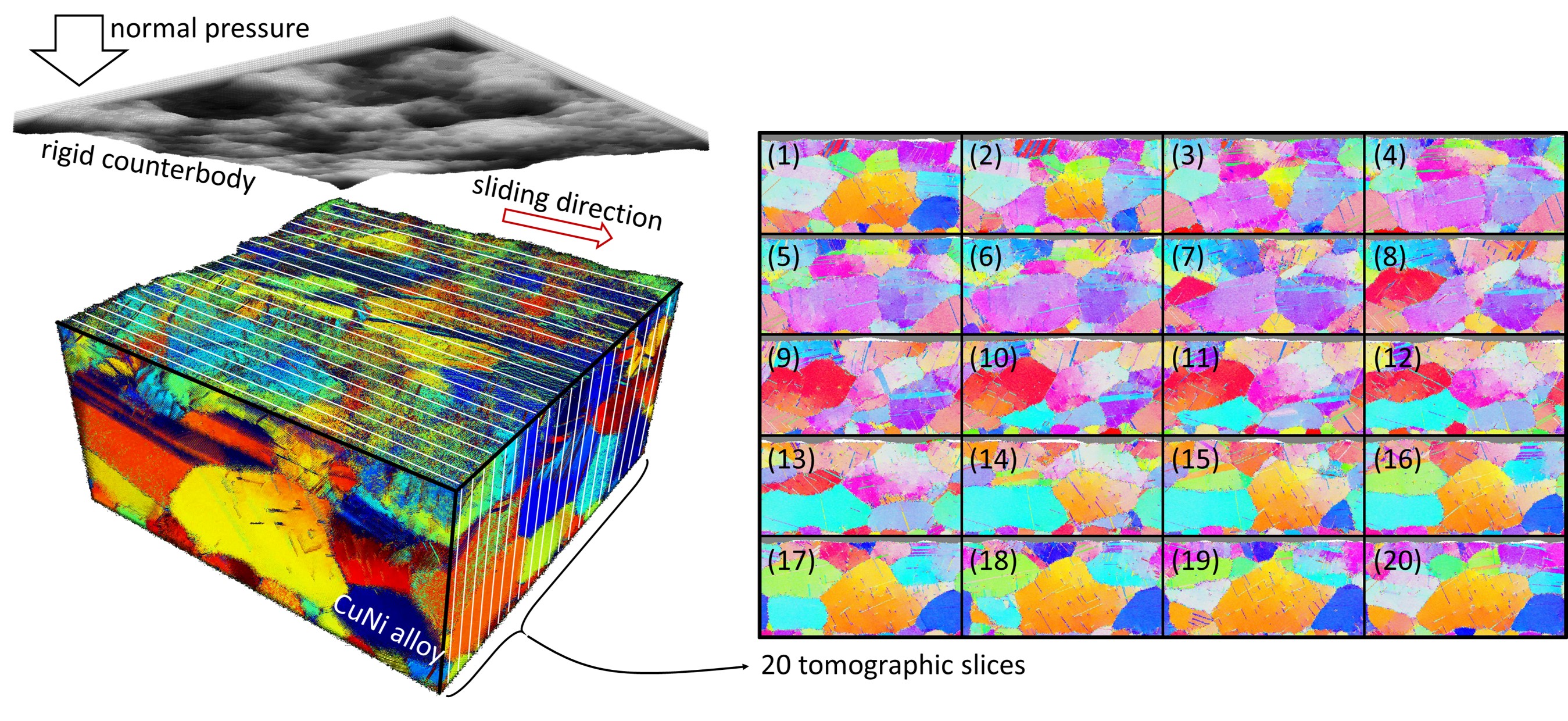}
    \caption{3D MD system setup overview sketching out the location of the tomographic slices that form the image basis for the ML training data.}
    \label{fig:SysSetup}
\end{figure}

The dataset used in this work consists of CuNi alloy simulation snapshots (Fig.~\ref{fig:SysSetup}) with varying Ni content (0\%, 5\%, 25\%, 60\%, and 100\%), normal pressure (ranging from 0.1~GPa to 2.7~GPa depending on composition, see Fig.~\ref{fig:SimCondsOverRegimes}~a), with constant temperature (300~K) and sliding speed (80~m/s).
Each system is represented as a series of computational tomographs sliced normal to the $y$-direction (with $x$ being the direction of sliding and the sliding surface being normal to the $z$ direction), each carrying associated metadata (Ni content, normal pressure, temperature, time step, and slice position).

\subsection{Labels}

\begin{figure}[htbp]
    \centering
    \includegraphics[width=0.99\linewidth]{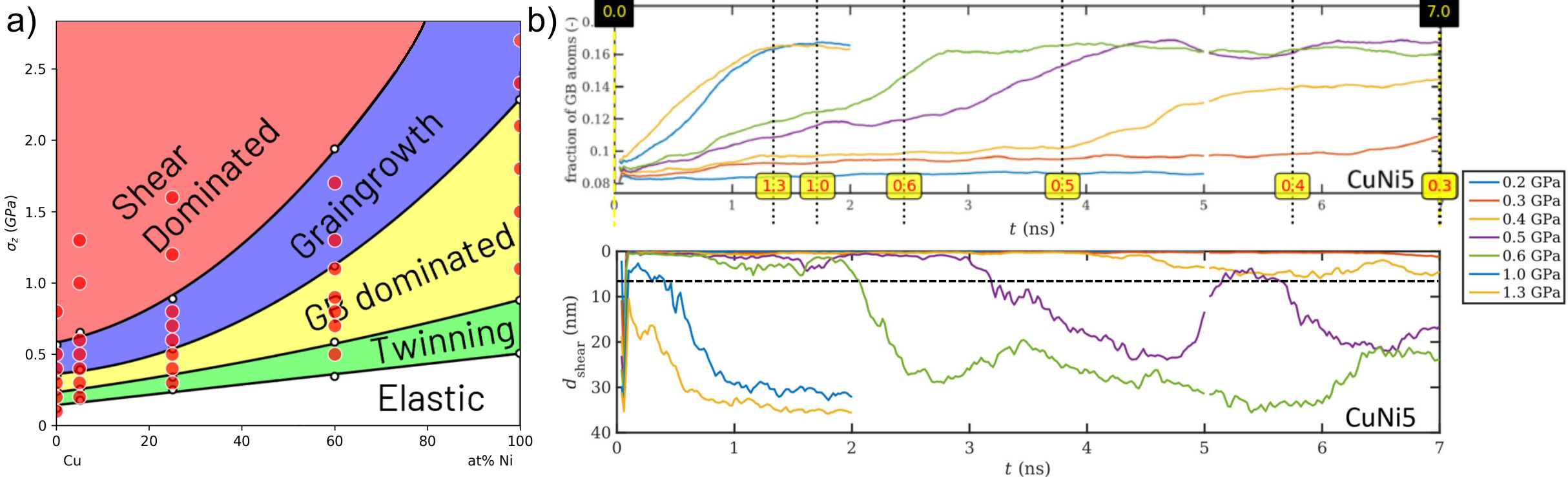}
    \caption{a) Deformation pattern map from~\cite{eder_unraveling_2020} as a function of Ni fraction and normal pressure, with the deformation regimes that serve as labels for the final states and the parameter combinations of the MD simulations marked as red circles. b) Exemplary intermediate plots of relevant time-dependent quantities (top: grain boundary fraction, transition times labeled; bottom: shear layer depth, no distinct transition times) used to assign labels to the transient states.}
    \label{fig:SimCondsOverRegimes}
\end{figure}

Initially, labels for each simulation run were derived directly from the system conditions, specifically alloy composition and applied normal pressure, which define the axes of the deformation pattern map previously established through manual analysis~\cite{eder_unraveling_2020}.
In this approach, each simulation corresponded to a single point in the deformation pattern map, such as shear-dominated or elastic, resulting in a straightforward labeling procedure of the final outcome of a single simulation run (see Fig.~\ref{fig:SimCondsOverRegimes}~(a)).
However, this method underutilized the available data, as each label encompassed multiple simulations with thousands of time-resolved samples, many of which corresponded to early stages of the simulation in which the microstructure did not yet resemble the final deformation pattern.
Consequently, the image data associated with each label was highly heterogeneous, and first models trained with these labels tended to rely predominantly on metadata rather than visual information.

To address this limitation, transient state labels derived from interim results of the manual workflow were incorporated.
These intermediate metrics, such as the evolution of the shear layer depth or the twin boundary and stacking fault fraction over time, provide insight into the current structural state of the simulation rather than only the final outcome.
Transient labels were assigned based on thresholds in these metrics (see Fig.~\ref{fig:SimCondsOverRegimes} b) top).
The available data supported reliable labeling only for the elastic, twinning, grain-boundary–dominated, and shear-dominated regimes, which is why the transient-state classifier was restricted to these four classes.
This approach enables the model to associate images with the structural state they depict, ensuring that visual information is meaningfully incorporated into training and improving the model’s capacity to capture the temporal evolution of deformation patterns.

Many of the metrics used to derive transient states, such as twin boundary and grain boundary fractions, are often correlated, meaning that a single time threshold may be insufficient to capture the true evolution of the deformation pattern.
Furthermore, the labeling process does not account for boundary states, where the microstructure exhibits characteristics of multiple regimes simultaneously (see Fig.~\ref{fig:SimCondsOverRegimes} b)).
To address these issues, transient labels could be subdivided based on the evolution of key quantities, such as shear layer depth, and assigned throughout the simulation in a time-resolved manner.
This assignment should also take the final state into account, ensuring physically consistent labeling (e.g., an ultimately elastic simulation would not include a shear-dominated transient state at any point).
This approach enables a more accurate and consistent representation of the evolving structural state in the images, and is pursued in ongoing work.
% improving the relevance of visual information for model training.

\subsection{Dataset Curation}
\label{subsec:DatasetCuration}
The available simulation data provided nearly 200,000 tomograph--metadata samples.
However, due to imbalances originating from both the simulation design and the structure of the labeling scheme, only a balanced subset of approximately 12,000 samples can ultimately be used for model training and evaluation (e.g., only one simulation was explicitly situated in the elastic regime, so a lot less data was available for this scenario).
This balancing ensures that all transient-–final label combinations were represented sufficiently often, which is essential because the goal of the model is to learn characteristic microstructural signatures, not the arbitrary frequency of deformation patterns in the simulation campaign.
Balancing joint labels in this way, however, introduces a secondary effect.
When the transient and final labels are considered independently, their individual distributions are no longer balanced.
Some classes naturally occur in many more transient–-final combinations than others, so the balancing by joint labels implicitly oversamples classes with fewer combinations and undersamples those with many.

The practical relevance of this imbalance is limited, but worth noting.
Since the learning target is the class-specific structure, not class frequency, the model does not use marginal frequency as a cue for prediction, and balancing does not distort the physical interpretation of the classes.
The model is thus exposed to all classes sufficiently often, which is the intended effect.
The main consequence is confined to evaluation rather than training, in that per-class accuracies for transient and final labels may not reflect the class proportions in the full simulation dataset.
Instead, they reflect the intentionally equalized joint-label sampling strategy.

The simulation campaign itself did not sample all compositions under loading conditions that would lead to every deformation pattern at least once.
For example, an elastic final state was only obtained for simulations of pure copper (see Fig.~\ref{fig:SimCondsOverRegimes}) for a visualization of the simulation conditions), whereas other alloy compositions consistently evolved toward shear-dominated or grain-boundary–dominated behavior.
This imbalance does not affect the image input branch, because atomic species are not visually distinguishable in the tomographic representations.
It does, however, introduce bias in the metadata branch, as certain metadata combinations, such as pure copper under low pressure, serve as strong indicators of specific final states, in this case the elastic regime.
Transient-state labeling is less influenced by these simulation-design constraints, since system evolution is tracked meticulously, and sufficient transient samples were available for each class.

\subsection{Unsupervised Representation Learning}
\label{subsec:unsupervisedMethods}

Autoencoders are ML models trained to reconstruct the input as accurately as possible by learning an encoder-decoder mapping \cite{yeh2017learning}. Concretely, an autoencoder consists of an encoder $f_\theta (x) = y$ that maps high dimensional, original inputs $x$ to a typically lower dimensional latent output $y$ and a decoder $g_\delta (y) = \hat{x}$. Given the model structures $f$ and $g$, both parameter sets $\theta$ and $\delta$ are adapted to reduce the reconstruction error $\mathcal{L} = ||x-\hat{x}||$.
Specifically, in an undercomplete autoencoder, this mapping $f$ compresses the input into a latent representation with a much lower dimensionality than the original data, in short $|x| >> |y|$~\cite{DeepLearning_Autoencoders}.
Thus, in this case in order to investigate the compressibility of high-dimensional MD tomographic images and assess whether essential structural information could be retained, we employed an undercomplete autoencoder with a 32-dimensional bottleneck layer using the mean squared error (MSE) as the loss function, ensuring that omissions of small but relevant defects were penalized during training.
Unlike most experimental machine-learning workflows for deformation pattern classification, where defects such as twins or stacking faults are manually annotated~\cite{SIEBERG2023204725}, the present approach aims for the model to discover and utilize relevant structural features autonomously.

\begin{figure}[h]
    \centering
    \includegraphics[width=0.9\linewidth]{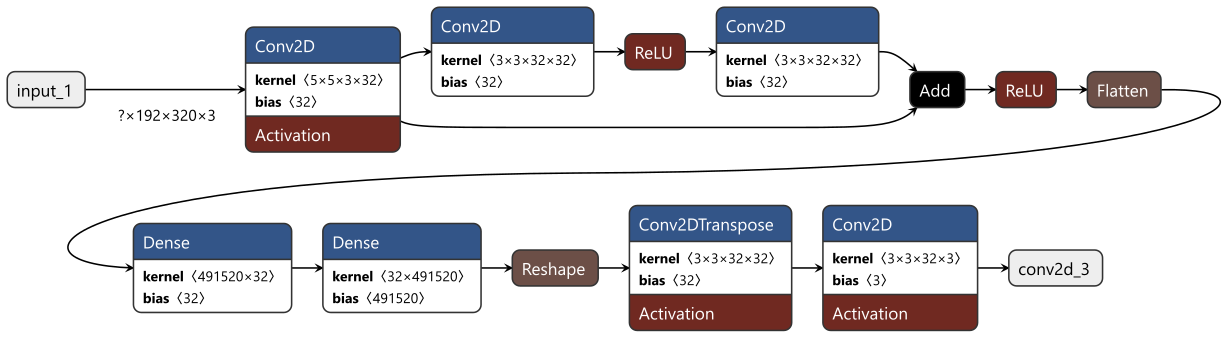}
    \caption{Undercomplete autoencoder with a shallow encoder–decoder design. A $5\times5$ stride-2 convolution and a residual block extract features and reduce spatial resolution before flattening into a 32-dimensional latent space. The decoder expands the latent vector via a dense projection, reshaping, and a stride-2 transposed convolution, followed by a final $3\times3$ sigmoid layer to reconstruct the image.}
    \label{fig:autoencoderArch}
\end{figure}

This approach aimed at reducing the images to a compact latent representation without requiring manual annotation of features such as twin boundaries, stacking faults, or phase transformations, as is often necessary in experimental transmission electron microscopy (TEM) or scanning electron microscopy (SEM) datasets.
In preliminary experiments, deeper architectures were evaluated, but produced visibly blurrier reconstructions, leading to the loss of fine structural features. As a result, we adopted a shallower architecture as visualized in Fig.~\ref{fig:autoencoderArch}.
Although this design led to a higher parameter count and a more abrupt transition from spatial representations to the flattened bottleneck, it preserved important local features more effectively.

Ultimately, the autoencoder was not employed as a preprocessing step for the subsequent architectures, as it has not yet been fully validated and its suitability for generating robust latent representations remains uncertain.

\subsection{Supervised Prediction Model}

\begin{figure}[htbp]
    \centering
    \includegraphics[width=0.6\linewidth]{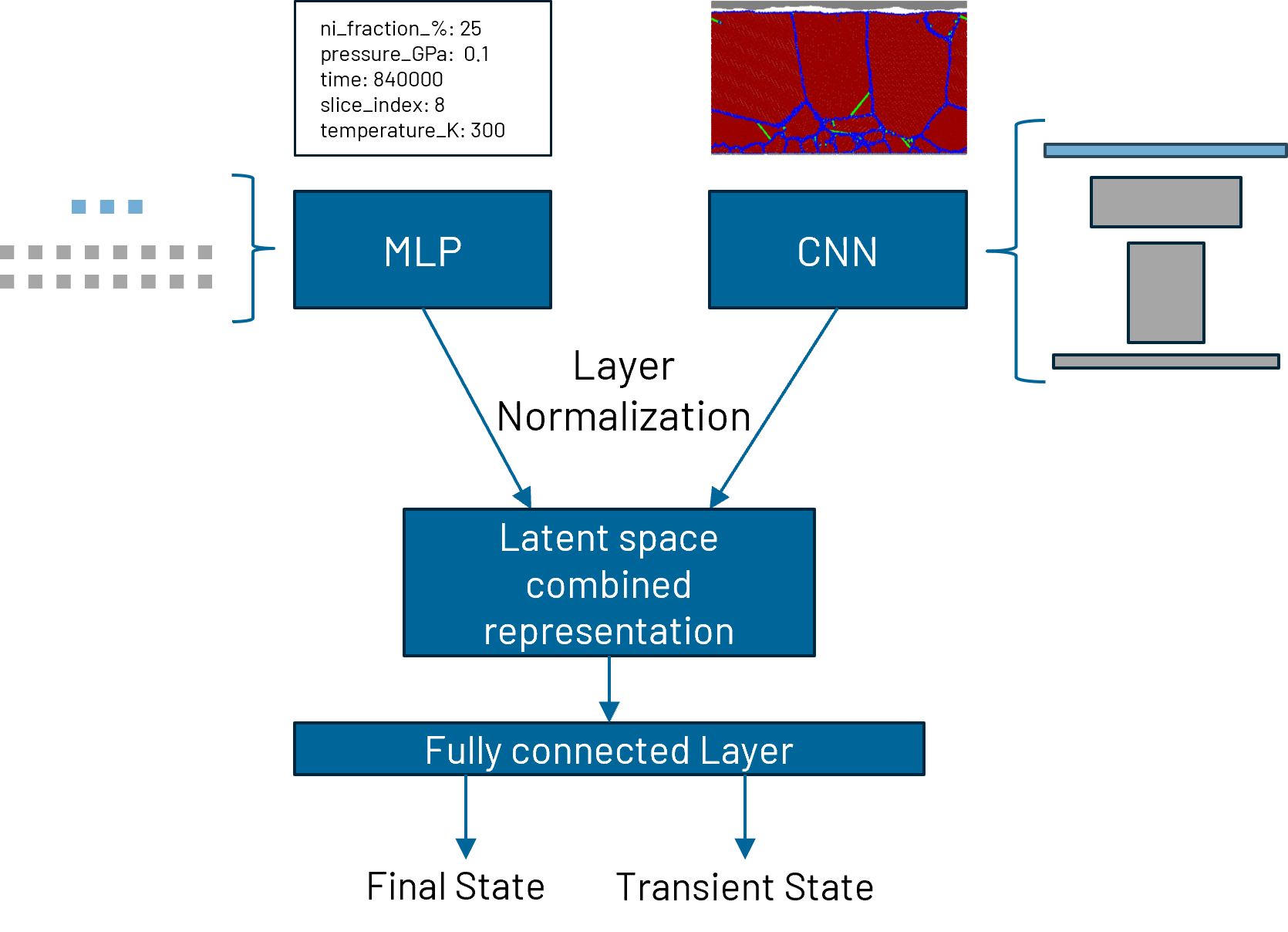}
    \caption{Two-branch neural network architecture combining numeric and categorical metadata and image data to predict two labels (4-class and 5-class). Image data is processed through a convolutional neural network, while metadata is handled via a Multi-Layer Perceptron (MLP). The outputs of both branches are concatenated and passed through additional fully connected layers to produce probabilistic predictions for both labels.}
    \label{fig:dualbranchnn}
\end{figure}

Independent of previously explored architectures, a dual-branch neural network \cite{gao2017dual}, consisting of two branches for image and metadata, was trained using paired transient and final state labels.
The architecture separately processes image data and the simulation metadata, in numerical form, before combining both modalities in a shared latent space as visualized in Fig.~\ref{fig:dualbranchnn}.

\begin{figure}
    \centering
    \includegraphics[width=0.9\linewidth]{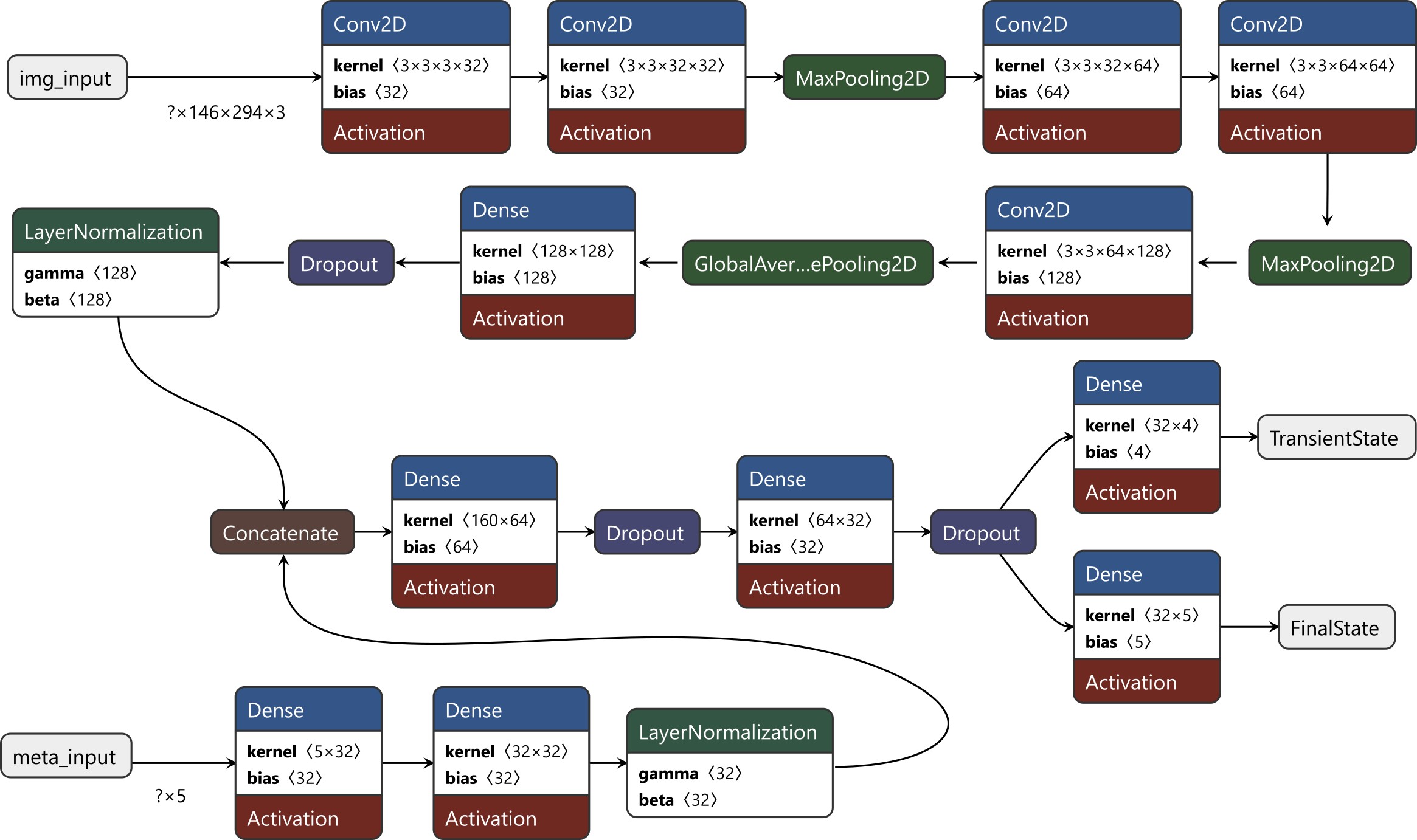}
    \caption{Detailed dual branch neural network architecture. The image branch (top) applies successive convolutional and dropout layers before flattening and projecting the features to a dimensionality comparable to that of the metadata branch (bottom), which consists of fully connected layers. The two feature streams are concatenated into a shared representation that is processed by additional fully connected layers to jointly predict the transient and final deformation labels.}
    \label{fig:nnarchitecture}
\end{figure}

More specifically, the concrete architecture was selected following the state-of-the-art approach of \cite{hunter2012selection} as dual branch neural network architecture optimization are notoriously hard to optimize with automated search architectures or AutoML \cite{luo2018neural}, especially if different modalities are regarded. 
The resulting architecture is visualized in Fig.~\ref{fig:nnarchitecture}.

The image branch consists of three convolutional blocks with 32, 64, and 128 filters, each using $3\times3$ kernels and ReLU activations, followed by max-pooling layers.
To reduce computational cost while preserving structural information, the input images were downscaled by a factor of five, keeping the original aspect ratio, resulting in an input size of $294\times146$ pixels. 
% Alternatively, the autoencoder compression is applicable to achieve the same input size.
A global average pooling layer preserves spatial information while drastically reducing the number of parameters, producing a feature vector that passes through a 128-unit dense layer with ReLU activation, dropout (0.3), and layer normalization.

On the other hand, the metadata branch comprises two fully connected layers of 32 units each, also followed by layer normalization.
At this stage, metadata is sparse as only a single simulation configuration is available, varying pressure and alloy composition (Ni fraction), but with constant temperature and sliding speed. 
As a result the corresponding weights are nilled, the overall architecture however shall be applicable to various simulation configurations, each with varying temperature and sliding speed.
Thus, the metadata inputs consist of Ni fraction, normal pressure, timestep, temperature, and slice index (analogous to spatial ($y$-)position within the system).

Outputs from both branches are concatenated and passed through two additional dense layers (64 and 32 units, ReLU activations) with dropout regularization following best practice \cite{hunter2012selection}.
The network produces two outputs, one, via a five-class softmax for the final deformation state and the other one via a four-class softmax for the transient state. 
Softmax is applied due to superior classification capabilities in earlier studies \cite{liang2017soft}.
The resulting model contains approximately 170k parameters, largely due to the parameter reduction achieved by the global average pooling layer.
Training was conducted for 60 epochs with a batch size of 32, using the Adam optimizer~\cite{kingma2014adam} and categorical cross-entropy as the loss function \cite{feng2021can}.

\subsection{Revised Validation Strategy}
\label{subsec:TrainValSplits}
Neighboring tomographs exhibited strong microstructural similarity due to their spatial proximity within the simulated volume.
Consequently, a random train–validation split would result in both sets containing highly similar microstructures, leading to artificially inflated evaluation metrics.
\begin{figure}[htbp]
    \centering
    \includegraphics[width=0.99\linewidth]{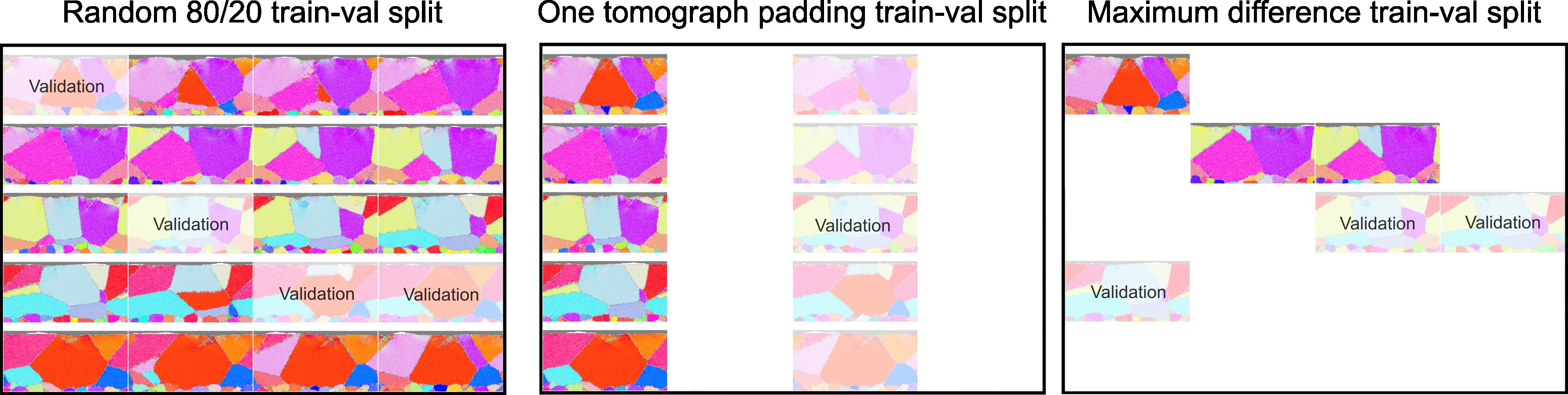}
    \caption{Training--validation split strategies. The 20 tomographs representing the 3D CuNi model are arranged in rows and periodic (bottom right end feeds back into the top left). (a) Classic random splitting of the data into 80\% for training and 20\% for validation. (b) ``One tomograph padding" strategy, where 50\% of the data is selectively discarded to reduce self-similarity, and the validation dataset is interleaved between the training set with only little overlap by similarity. (c) Maximum difference split, where 70\% of data was discarded and tomographs for training and validation were selected manually so that there is the least possible overlap (worst-case scenario).}
    \label{fig:TrainValSplits}
\end{figure}
In such a case, the model would effectively have been exposed to each microstructural configuration at least once, even if at a different simulation time, thereby testing recognition rather than generalization.
To more rigorously assess the model’s ability to generalize, contiguous tomographic regions were excluded from training.
This strategy mimics evaluation on unseen microstructures, similar to testing on newly generated simulations with independently initialized microstructures.
Strong performance under these conditions indicates that the model learned generalizable representations of microstructural features rather than memorizing specific image patterns \cite{yu2015lsun}.

For this purpose, the five most distinct tomographs from the dataset were selected for training, with one tomograph on either side reserved as validation data (see Fig.~\ref{fig:TrainValSplits} b)) to provide a controlled but structurally distinct test set.
To stress-test the model, we also performed an extreme experiment in which the model was trained on only three tomographs (two maximally different positions plus one additional slice adjacent to one of those positions, see Fig.~\ref{fig:TrainValSplits} c)) and validated on tomographs maximally different from the training data.

\subsection{Model Interpretation}
Model interpretability is achieved using SHAP (SHapley Additive exPlanations)~\cite{lundberg2017SHAP}, which attributes a quantitative contribution to each input feature for a given prediction.
SHAP is based on Shapley values from cooperative game theory, where each feature is treated as a ``player" contributing information to the model’s output.

For a target sample, SHAP evaluates the model repeatedly on combinations of feature subsets constructed from representative background data.
Each feature’s true value is paired with background values for the remaining features, and the resulting change in model prediction is compared to the case where that feature is replaced by background values.
Averaging these prediction differences across many feature subsets yields the SHAP value for that feature, representing its average marginal contribution to the prediction relative to the dataset baseline \cite{lundberg2017SHAP}.

The model output for each sample can thus be expressed as
\begin{equation*}
    f(x)=E[f(X)]+\sum_j\phi_j(x),
\end{equation*}
where $E[f(X)]$ is the expected model output over the dataset, and $\phi_j(x)$ is the SHAP value quantifying the contribution of feature $j$.
Positive SHAP values indicate that a feature increases the model’s evidence toward a specific class, while negative values indicate the opposite.
When computed on the model’s logit outputs, these contributions are additive and directly comparable across classes \cite{wang2024feature}.

\section{Results}
\subsection{Feature Representation and Reconstruction}
The autoencoder architecture introduced in Section~\ref{subsec:unsupervisedMethods} was trained on two types of tomographic images.
Electron backscatter diffraction (EBSD)-colored tomographs encoding grain orientation on the one hand, and Common Neighbor Analysis (CNA)-colored tomographs, which encode local structural classifications (fcc, hcp, bcc, undefined and areas with advection velocity in x direction of over 10 $m/s$), on the other.
These representations emphasize microstructural features differently and therefore probe different aspects of compressibility.

\begin{figure}
    \centering
    \includegraphics[width=0.7\linewidth]{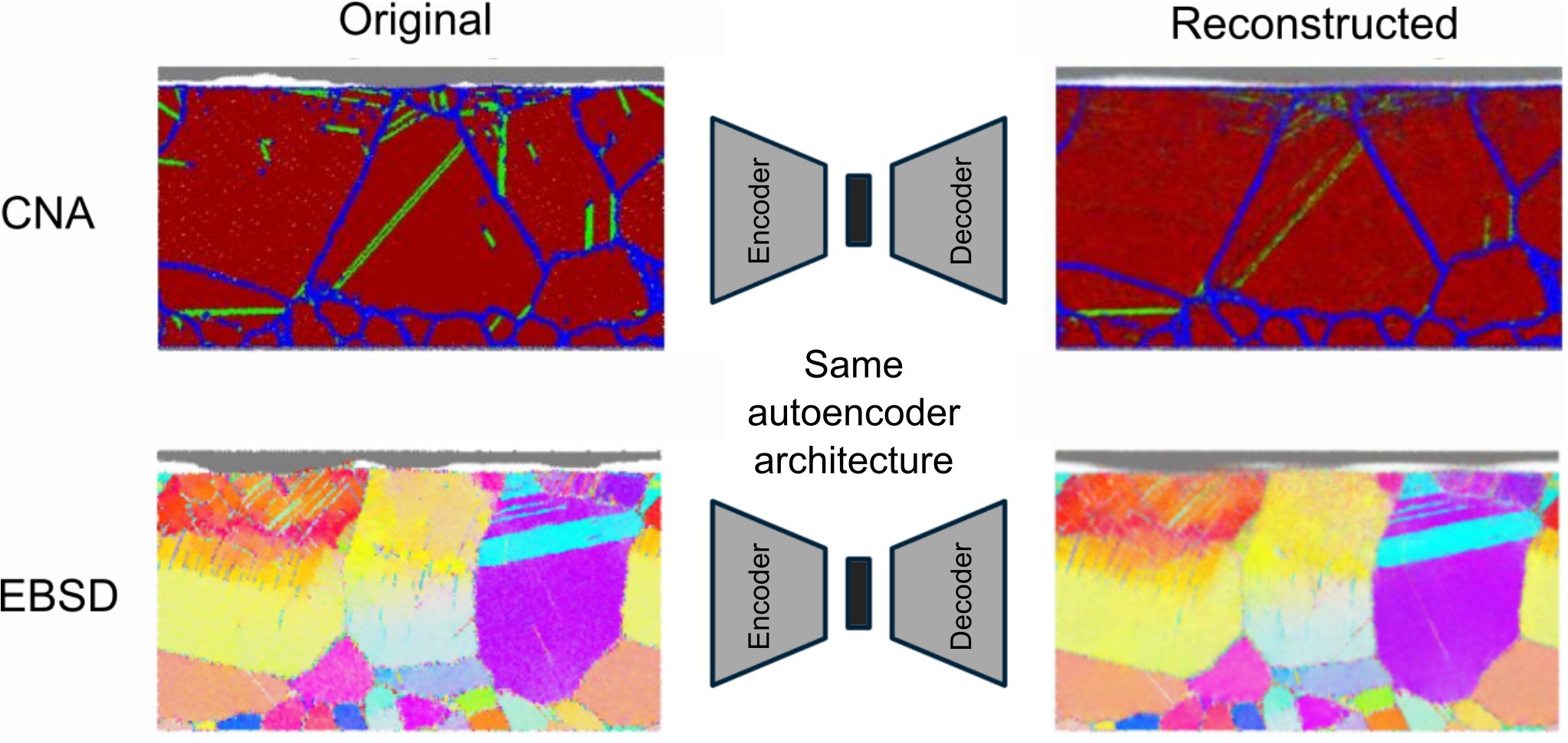}
    \caption{Autoencoder reconstructions of MD tomographs highlighting microstructure (CNA) and grain orientation (EBSD). The rigid counterbody is shown in gray. Separate models were trained on each representation to compare reconstruction performance. For CNA representations, grain boundaries are accurately preserved, while small-scale defects like stacking faults are often lost during reconstruction. Dense defect regions show reduced fidelity.
    EBSD images retain fine structural details.}
    \label{fig:autoencoderresults}
\end{figure}

In the CNA-colored images, the dominant features are grain boundaries and dislocations (in blue), and extended defects such as stacking faults and twins (in green), which appear as sharp, localized deviations within otherwise uniformly colored crystalline grains (in red).
The autoencoder reconstructed grain boundaries reliably as evidenced in Fig.~\ref{fig:autoencoderresults}, indicating that the global morphology of the microstructure was well captured in the low-dimensional latent space.
However, isolated defects such as single stacking faults were frequently omitted in the reconstructions.
This behavior was reflected directly in the training curves as the MSE loss plateaued after a moderate number of epochs, consistent with the pixel-wise loss function penalizing small, spatially sparse defects that the network could not reproduce faithfully.
Because the loss is dominated by the large, homogeneous grain interiors, these small defects had negligible influence on optimization and remained unreconstructed.

By contrast, the EBSD-colored tomographs, despite visually containing more complex color variations and similarly fine details, were reconstructed with much higher accuracy.
Even localized defects, including individual stacking faults, were retained in the reconstructed images.
This difference indicates that the orientation-based representation distributes microstructural information more smoothly across the image domain, enabling the autoencoder to encode subtle rotational variations and defect signatures more effectively within the limited available latent dimensionality.

Although the autoencoder confirmed that the tomographic data can be compressed to a low-dimensional representation, the simplicity of the final architecture imposes constraints on the generality of the learned latent space.

The encoder employs a shallow architecture with few convolutional operations before projection into the latent space.
This design performs an aggressive dimensionality reduction after only minimal hierarchical feature extraction.
Consequently, the latent variables predominantly reflect the specific filtering characteristics of this shallow encoder rather than a broadly structured feature hierarchy.
As previously mentioned, deeper encoder variants with additional convolutional stages were also evaluated, but these models did not achieve comparable reconstruction performance.
Their training usually failed to converge towards a reconstruction quality that preserved the relevant microstructural detail.
This outcome indicates that the apparent architectural narrowness of the final model is not merely a design choice, but a practical constraint imposed by the data characteristics, the limited variability in the training set, and the pixel-wise MSE objective, which harshly penalizes even small misalignments.

As a result, the learned latent space should be interpreted primarily as a task-specific compression tailored to this particular architecture and reconstruction objective, rather than as a general-purpose representation suitable for downstream transfer.
Because the flattening step collapses spatial organization abruptly, the model has no mechanism to form intermediate, physically interpretable feature hierarchies that would support cross-task reuse.

\subsection{Prediction Performance}
The multi-modal deformation pattern classification model was trained using the three training–validation splits introduced in Section \ref{subsec:TrainValSplits}.
The observations summarized below hold for all splits, with differences only in degree.
Performance variations arising from the specific dataset configurations are examined in the following section.
The trained dual-branch model demonstrated strong generalization performance with no indications of overfitting.
Prediction accuracy for final deformation states exceeded that for transient states.
This is expected, as the labels (deformation pattern defined by the Ni fraction and pressure) are constant for each simulation, so the combination of final microstructure and consistent metadata appears frequently and is therefore easier for the model to learn.
By contrast, transient states exhibit greater structural variability and weaker coupling to the metadata, making them inherently more difficult to classify.

\begin{figure}[htbp]
    \centering
     \includegraphics[width=0.99\linewidth]{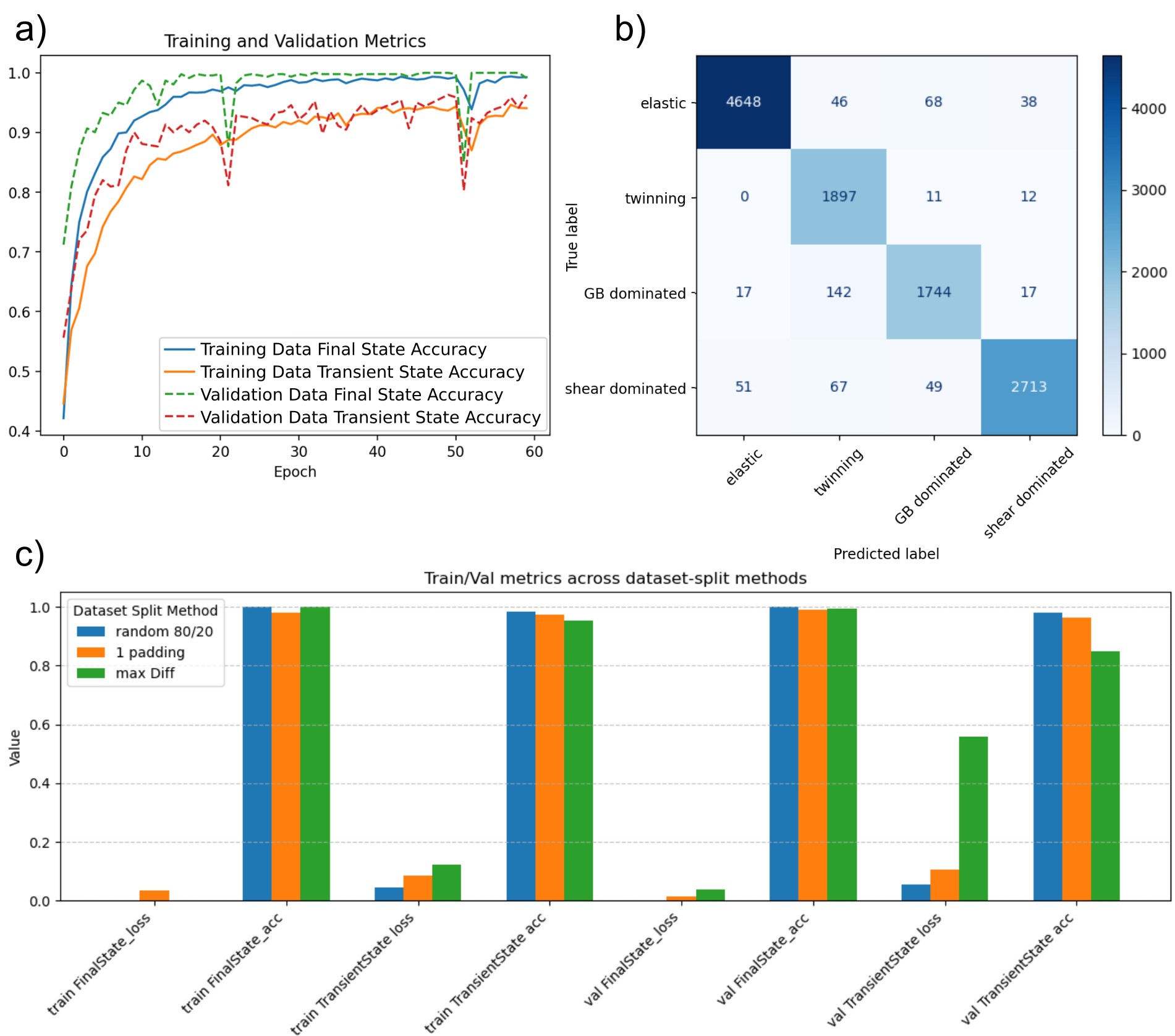}
     \caption{Model training and performance metrics. (a) Accuracies of training and validation data for final and transient state predictions during training with the ``1 padding" train--val split. (b) Confusion matrix for true and predicted labels for the transient states (``1 padding" train--val split). (c) Comparison of losses and accuracies between the three different train--val split approaches.}
    \label{fig:PredPerf}
\end{figure}

Validation accuracy occasionally exceeded training accuracy due to dropout being active only during training.
When the model was re-evaluated on both training and validation datasets under identical inference conditions (dropout disabled), their performance was nearly identical, confirming that the network is not overfitting and that the apparent discrepancy arises solely from the regularization behavior of dropout (see Fig.~\ref{fig:PredPerf}) \cite{eelbode2021pitfalls}.

\subsection{Generalization Across Spatial Regions}
Using the spatially independent validation strategy introduced in Section~\ref{subsec:TrainValSplits}, the model’s ability to generalize beyond microstructures it had been exposed to during training was tested.
Because of the grain sizes and periodic boundaries in our simulations, it is nearly impossible to find tomographs that do not share grains with others.
However, shifting roughly five slices left or right typically produces a visibly different scene with distinct grain morphologies satisfying the suggestions for human in the loop for image based machine learning \cite{yu2015lsun}.
Although this approach naturally reduces the apparent accuracy compared to a random train–validation split, since large contiguous regions of similar tomographs are withheld, it provides a more realistic measure of true generalization.
Even under these stricter conditions, the model achieved $\sim$99\% accuracy for final-state prediction and >95\% for transient-state prediction, demonstrating that it learned robust, transferable representations rather than relying on microstructural similarity between neighboring slices.
Even in the highly constrained setting of only 3 images for training (``maxDiff"), the model retained strong predictive power, achieving $\sim$95\% accuracy for final-state predictions and $\sim$85\% accuracy for transient-state predictions on the validation set.
These results indicate that the network learns transferable microstructural representations and can generalize to structurally distinct regions even under reduced training data.
Because excluding contiguous regions substantially reduces the amount of available training data, the model is effectively evaluated under information-limited conditions, yet still performs comparably to random-split experiments.

\section{Discussion}

The relative importance of image and metadata inputs was evaluated through an ablation study.
For final-state predictions, accuracy decreased sharply when only image data were used, whereas removal of the image branch had minimal effect, highlighting the dominant role of metadata for final state prediction.
For transient-state predictions, removal of either images or metadata resulted in an approximately 40\% decrease in accuracy, indicating comparable contributions from both input types.
These results demonstrate that predictions of transient deformation evolution rely on both structural information presented in the form of images and contextual information from metadata.

When interpreting these behaviors, it is important to note again that the model learns from \emph{paired} transient--final labels, since both outputs are predicted jointly.
This pairing inherently introduces uneven sampling across the combined label space.
% Nonetheless, pairing the transient and final labels, which is necessary because the model predicts both jointly, introduced additional imbalance.
Rare combinations, such as a grain-growth final state paired with a transient twinning state (a configuration that may occur only briefly during evolution), became strongly underrepresented.
Ideally, the MD simulation matrix would be expanded to ensure full coverage and balanced occurrence of all deformation patterns across compositions and loading conditions.
However, achieving such balance is the subject of future work, as the required simulations are computationally intensive and beyond the scope of the present study.

\begin{figure}[htbp]
    \centering
    \includegraphics[width=0.99\linewidth]{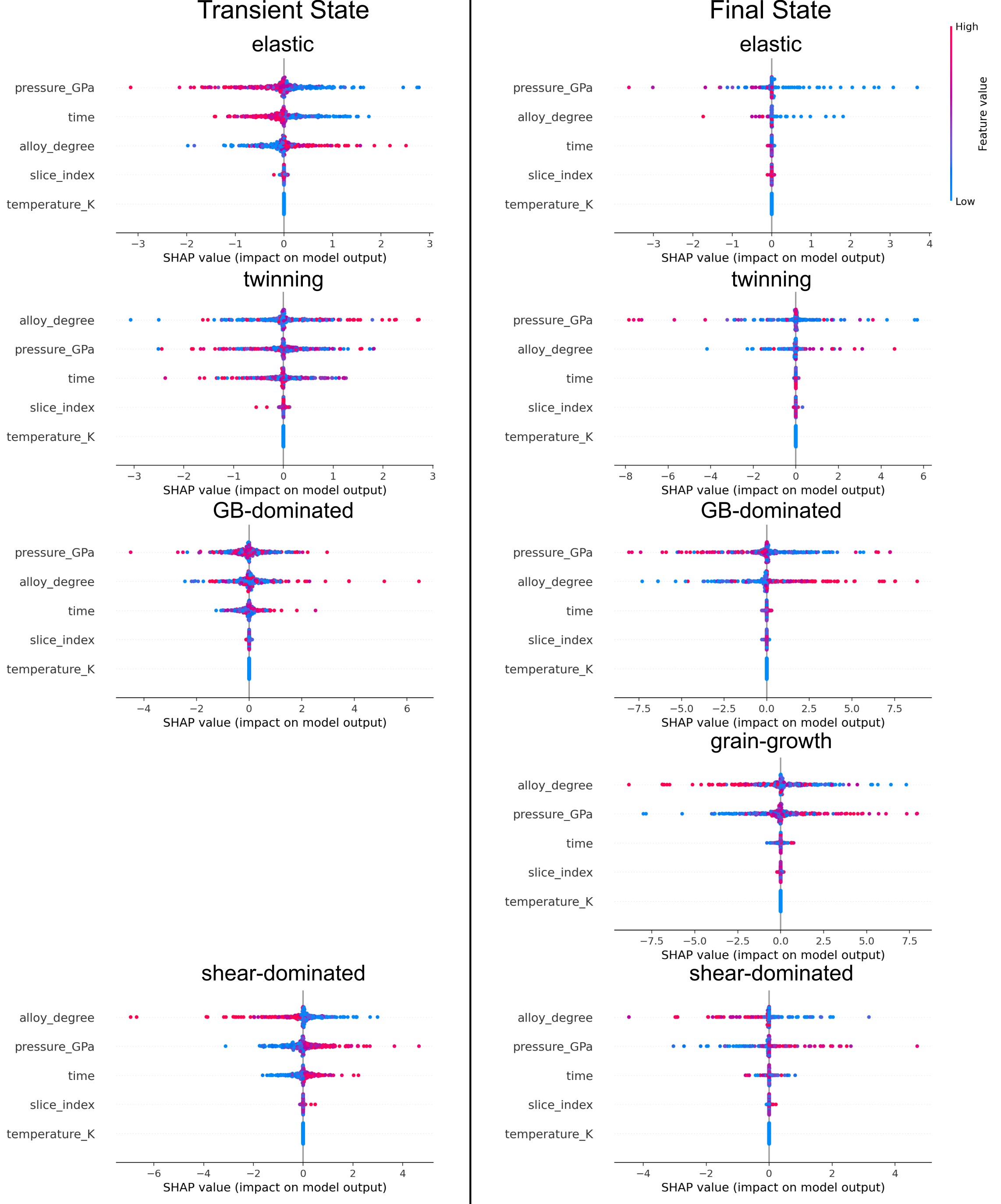}
    \caption{Beeswarm plots of the metadata SHAP analysis. Left column: transient state results, right column: final state results. See the main text for a detailed discussion.}
    \label{fig:ShapMetadata}
\end{figure}

Metadata SHAP values for each parameter and each class are visualized using class-specific beeswarm plots (see Figure~\ref{fig:ShapMetadata}).
For the transient classes, the extreme deformation states, purely elastic and fully shear-dominated, showed clear, interpretable trends.
Low pressure and high Ni fraction increased the probability of the elastic transient state, while progression in time decreased it as other deformation patterns emerged.
Conversely, high pressure, low Ni fraction, and later time stamps consistently increased the probability of the shear-dominated transient state.
These trends align with the labeling system used to assign the ground-truth labels.
The intermediate transient states, twinning and grain-boundary dominated, showed substantially weaker and less consistent metadata influence.
Here, none of the metadata parameters exhibited a systematic monotonic effect, and overall SHAP magnitudes were smaller than in the elastic or shear-dominated cases.

Counterintuitively, the final-state predictions exhibited even less clear metadata influence in the SHAP analysis, despite ablation studies indicating that the model relied almost exclusively on metadata for these classifications.
This apparent contradiction can be explained by the labeling procedure.
Multiple combinations of pressure and Ni fraction can lead to the same final state, so no single metadata trend dominates within a class.
Hence, fractional changes as in the SHAP analysis have little effect.
For the elastic final state, low pressure generally increased class probability, while Ni-fraction effects were mixed but included clear low-Ni cases with higher elastic likelihood, which results from the condition imbalance mentioned in Section~\ref{subsec:DatasetCuration}.

For the twinning final state, metadata SHAP values are scattered, with both high and low pressure or Ni fraction producing positive or negative contributions.
GB-dominated and grain-growth states exhibited similar ambiguity, though broader trends matched the widening of their respective regions in the deformation pattern map.
Higher Ni fraction and lower pressure favored GB-dominated outcomes, whereas grain-growth was promoted by a lower Ni fraction and higher pressure.
The shear-dominated final state showed the clearest metadata trend, with strong preference for low Ni fraction and high pressure.

The positional encoding (slice index) contributed negligibly across all classes.
Additionally, many samples exhibited SHAP values clustered near zero for individual metadata entries, suggesting that for a substantial subset of cases no single metadata parameter had a decisive marginal effect on the model output.
This reflects the multivariate nature of the pattern map, where classifications may depend on combinations rather than isolated metadata variables.

\begin{figure}
    \centering
    \includegraphics[width=0.99\linewidth]{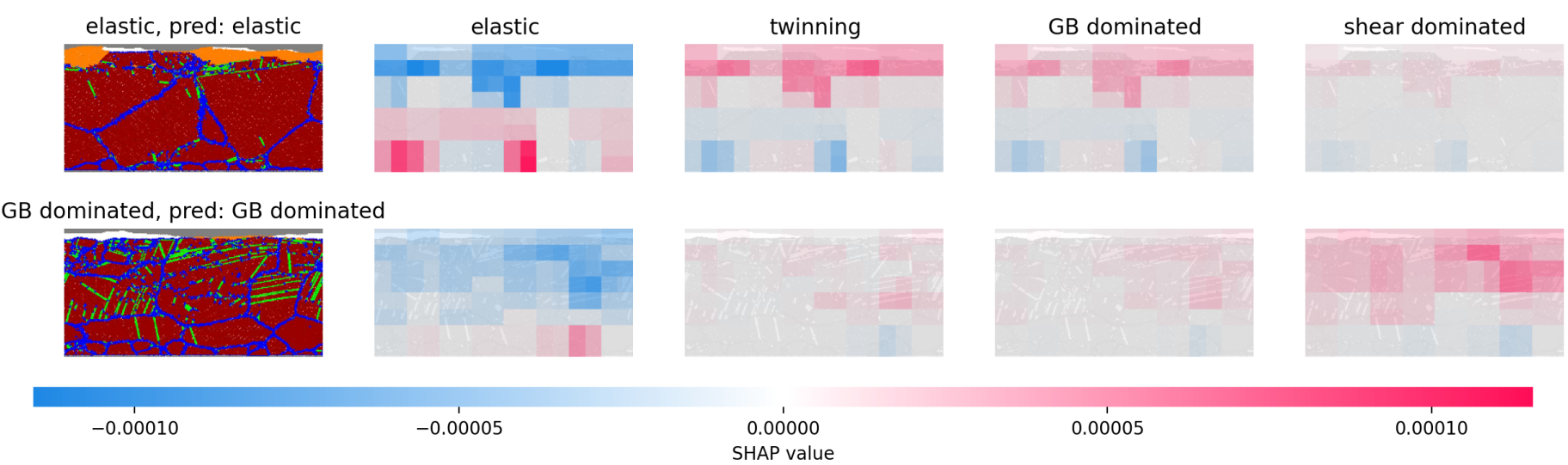}
    \caption{Image branch SHAP analysis. Two representative examples highlighting how SHAP may help explain how an ML model came to the conclusion to predict a given label. Image overlays show regions of identified positive (red) and negative (blue) class prediction influence. Top row: while the top end of the image (with the sliding interface) would favor a prediction of either twinning or grain-boundary (GB) dominated deformation patterns, the lower end favors the elastic regime. Bottom row: The visible grain refinement and defect density in the image on the left correctly precludes a labeling as elastic and suggests shear-dominated behavior, but consideration of the metadata leads to labeling as GB-dominated.}
    \label{fig:shapImgExp}
\end{figure}

For the image branch, SHAP was applied using partition-based explanations, which decompose an image into smaller regions and estimate each region’s contribution to the class probability.
The resulting overlays color-code positive and negative influence, allowing spatial interpretation of where the model ``looks” when assigning a label.
Unlike metadata beeswarm plots, these explanations cannot be aggregated into a meaningful class-level average, as they represent SHAP values for a single sample only, with a different dimensionality.
The computational cost is substantially higher than for metadata when perturbed versions of each image are evaluated against the background dataset.
Consequently, metadata must be held fixed during image SHAP computation, as allowing both image regions and metadata values to vary jointly would be prohibitive in dual branch cases \cite{wang2024feature}.

In practice, interpreting the image-based SHAP results is challenging.
For most samples, overlays intuitively highlight microstructural features, e.g., regions with strong deformation contributing negatively to the elastic class probability (Fig.~\ref{fig:shapImgExp}).
However, in many cases the SHAP-attributed influential regions do not align with the model’s actual predicted class.
% Based solely on the overlay, a different classification than the one produced by the model might be expected.
Even when evaluating metadata SHAP values and image SHAP values for the same sample, it is often difficult to derive a coherent explanation.
Additional effort may therefore be required to better understand and interpret these outputs, especially given the high-dimensional, multimodal nature of the model.

\section{Conclusions \& Outlook}

% More notes:
% \begin{itemize}
%     \item more than just classification of a current state - evolution, etc
%     \item new representations, new architectures
%     \item training with data from new MD runs -> multiple simulations
%     \item training with highly loaded systems that pass through several modes of deformation in a short period of time?
% \end{itemize}

% -----------------------------------------------
The workflow developed in this study establishes a first step toward automated tribological deformation mechanism mapping based on large-scale molecular dynamics simulations using Machine Learning.
By demonstrating that deformation patterns can be inferred directly from structural tomographs using learned feature representations, we provide a proof of concept for data-driven mechanism identification.
The dual-branch CNN–MLP model achieves high classification accuracy even on previously unseen configurations, and the latent-space analysis confirms that essential microstructural characteristics can be captured in a condensed, compact form.
Together, these results highlight the feasibility of automating deformation regime labeling and of building predictive tools that can augment, accelerate, or partially replace costly MD simulations and subsequent steps.

Neural network architectures offer a large search space for optimizing their performance, including architecture, convolutional depth, fusion strategies, activation functions, regularization schemes, and loss formulations as well as hyperparameters. A systematic exploration of these requires AutoML approaches and significant time and effort and, thus, is suggested for future work.
A deliberately conservative architecture, based on existing state-of-the-art dual branch neural networks architectures, was therefore chosen to establish feasibility and to analyze generalization behavior.
The strong performance obtained nevertheless indicates the potential of marrying Machine Learning, MD simulations and tribological studies. In particular, additional architectural tuning and hyperparameter optimization could further enhance robustness and accuracy.

Future developments will expand the scope of the workflow well beyond static deformation-state classification to cover multiple stages of tribosystem analysis.
A key objective is to learn microstructural evolution itself, incorporating temporal information and exploiting richer data representations.
This includes voxel-based encodings that retain the spatial relationship of the 3-dimensional system, point-cloud descriptions of defects, grain boundaries or orientations, and graph-based representations in which grains form nodes that are connected by grain-boundary edges.
Such representations enable the application of advanced architectures like transformers~\cite{gao2024exploring}, physics informed neural networks (PINNs)~\cite{hasan2024microstructure}, graph neural networks (GNNs), spatio-temporal GNNs~\cite{qin2024graingnn,fan2024accelerate}, and physics-embedded graph networks for learning microstructure evolution directly on topological graphs~\cite{xue2022physics}.
Such architectures will allow, for instance, a transition from state classification to recursive evolution or pattern prediction, in which the model iteratively predicts future microstructural states and behavior.

Besides algorithmic enhancement and developments, a broader and more diverse MD dataset will be essential for future work.
This includes new simulation runs across alloy classes, compositions, pressures, and temperature ranges, as well as highly loaded systems that pass through multiple deformation modes within short time intervals.
Such highly loaded configurations serve not only to enrich the range of observable mechanisms but also to evaluate whether short simulations containing compressed sequences of deformation events are sufficient for learning material behavior.
This could allow the model to infer long-term or steady-state responses from truncated trajectories, thereby reducing the need for fully resolved and computationally expensive MD runs.
In a similar vein, hybrid approaches that capitalize on execution in machine learning and are coupled with certain MD simulation aspects are worth exploring.

In the long term, the goal extends beyond deformation pattern detection from static tomographs.
The intended direction is towards a predictive system capable of inferring material behavior from minimal input, potentially from only initial microstructure, loading conditions, and published material properties, without requiring full MD trajectories.
Achieving this would move computational tribology closer to data-driven, mechanism-aware “tribological Ashby charts” and ultimately toward AI-accelerated simulation pipelines that guide or replace high-cost atomistic simulations in mechanism prediction and capture and formalize the knowledge contained therein.

\section*{Acknowledgement}
Part of the presented results were realized as part of the COMET Centre InTribology (FFG no. 906860), a project of the “Excellence Centre for Tribology” (AC2T research GmbH). 
COMET is managed by The Austrian Research Promotion Agency (FFG).
The computational results presented here were obtained using the Vienna Scientific Cluster (VSC).
% Open access funding was provided by TU Wien (TUW).

% \printcredits

% \section*{Declaration of Generative AI and AI-assisted technologies in the writing process}

% During the preparation of this work the authors used GPT-5 in order to shorten and improve the readability of the introduction. 
% After using this tool, the authors reviewed and edited the section as needed and take full responsibility for the content of the publication.

% \section*{Data Availability Statement}

% \FloatBarrier
%% Loading bibliography style file
% \bibliographystyle{model1-num-names}
\bibliographystyle{unsrt}

% Loading bibliography database
\bibliography{refs}

% \clearpage
% \setcounter{figure}{0}
% \appendix

% \section{Simulation and Model Details}
% Include detailed parameters of the MD setup (system dimensions, atom counts, potential, loading conditions) and neural network hyperparameters (architecture, optimizer, loss functions, etc.).

% \appendix
% \section{My Appendix}
% Appendix sections are coded under \verb+\appendix+.

% \verb+\printcredits+ command is used after appendix sections to list 
% author credit taxonomy contribution roles tagged using \verb+\credit+ 
% in frontmatter.

\end{document}